\documentclass[prl,showpacs,twocolumn,aps]{revtex4}
\usepackage{amssymb}

\begin{document}

\title{Sure success partial search}
\author{Byung-Soo Choi}
\email{bschoi@cs.york.ac.uk, bschoi3@gmail.com}
\author{Thomas A.\ Walker}
\author{Samuel L.\ Braunstein}
\affiliation{Department of Computer Science, University of York,
United Kingdom}
\date{\today}

\begin{abstract}
Partial search has been proposed recently for finding the target
block containing a target element with fewer queries than the full
Grover search algorithm which can locate the target precisely. Since
such partial searches will likely be used as subroutines for larger
algorithms their success rate is important. We propose a partial
search algorithm which achieves success with unit probability.
\end{abstract}

\maketitle

\section{Introduction}
In 1985 Deutsch designed a quantum algorithm which evaluates whether
the two outputs of a Boolean function are the same or not using only
one function evaluation \cite{CS-0900.81019MA}. Deutsch and Jozsa
generalized this algorithm for a more general case such as whether a
given Boolean function is constant or balanced. This algorithm
demonstrated an exponential speed-up on a quantum machine compared
to the best performance on classical machines \cite{qDJ92}. The most
important contribution in this field was achieved when Shor
discovered a polynomial-time quantum algorithm for factoring and
computing discrete logarithms --- yielding an exponentially faster
algorithm than the best known classical ones
\cite{SICOMP::Shor1997}. After this breakthrough many researchers
started to find various applications, especially in cryptanalysis.
On the other hand, Grover discovered the quantum (virtual) database
search algorithm which yields a quadratic speed-up compared to
classical database searches \cite{qGR96}. Since the database search
algorithm is one of the most widely used algorithms in computer
applications, the scientific impact is huge and many researchers
have been interested in various applications of the quantum database
search algorithm. The work presented here is concerned with a
variation of the Grover search algorithm.

Recently, several researchers have investigated a partial search
where instead of seeking the exact location of a unique target
solution, they are interested in finding which `target block' the
solution sits in \cite{GroverRadhakrishnan,KorepinGrover}. Indeed,
because only a partial search is being performed an improvement in speed
over the full search is expected. Indeed, recently proposed
algorithms achieve meaningful performance improvements over full
search \cite{Korepin}.

Meanwhile, until now these works have considered only optimizing the
performance at the expense of finding the target block with unit
probability. One might hope that partial search could become an
important component or subroutine of larger quantum algorithms if a
sure success (unit probability) formulation could be found. The idea
would be to perform the search on successively smaller block sizes
with each partial search successively revealing more information
about the location of the target. Indeed, in
Ref.~\onlinecite{GroverRadhakrishnan} the idea for a sure success
partial search has been mentioned. In order to achieve this goal we
utilized the scheme for partial search as described in
Ref.~\onlinecite{KorepinLiao} which reduces the problem to one
essentially involving rotations in a three-dimensional Hilbert
space. In this way we find that a simple modification, involving
introducing additional phases in the final step, allows us to construct
a sure success partial search algorithm.

This paper is organized as follows. Firstly, we review an optimal
version of the partial search algorithm, known as GRK algorithm
\cite{Korepin}. Secondly, we propose a modification to the phases
for the final step to guarantee sure success of the partial search
algorithm. We derive a phase condition that must be satisfied for
this modification to yield sure success and finally we show
numerically that this condition may easily be solved. We conclude
with a consideration of other problems that might be extensions
suitable for further study.

\section{GRK Partial Search Algorithm}
The GRK partial search algorithm \cite{Korepin} is defined by the
sequence of unitary operations
\begin{equation}
G_gG^{j_l}_lG^{j_g}_g\;,
\end{equation}
where $G_g$ is a `global' Grover operator which acts on the entire
search space of size $N$ and $G_l$ is a `local' Grover operator
which acts on the local search space in each non-overlapping block
$B_i$ of size $b$, $i=1,\ldots,K$. The initial state, uniformly
superposed over all $N$ input values, is
\begin{equation}
|\psi_{\rm
init}\rangle=\frac{1}{\sqrt{N}}\sum_{x=0}^{N-1}|x\rangle\;,
\label{init}
\end{equation}
and, similarly, the target state is a uniform superposition over the
single block $B_t$ containing the unique solution state in a regular
Grover search
\begin{equation}
|\psi_{\rm target}\rangle=\frac{1}{\sqrt{b}}\sum_{x\in
B_t}|x\rangle\;.
\label{target}
\end{equation}

Briefly, after $j_g$ iterations of the global operator $G_g$, the
amplitude of the solution state increases. Next, after $j_l$
iterations of the local operator $G_l$, the amplitude of the
solution state and the amplitudes of the remaining states in the
target block have been modified, with no change to the amplitudes of
all other states. Finally, one more iteration of the global search
operator $G_g$ is used. After these steps one would ideally want
non-zero amplitudes only those states within the target block.
Several combinations of the sequences of global and local operators
have been investigated and numerically the GRK formulation has been
found to be optimal (in number of calls to the oracle)
\cite{KorepinLiao}. For the purposes of this paper we shall assume
this result correct. Hence, we shall give more details about the GRK
construction \cite{KorepinLiao}.

Since the number of blocks $K$ and block size $b$ satisfy $N=Kb$ we
find it useful to define some trigonometric counterparts to them via
\begin{equation}
\sin^2\theta_g=\frac{1}{N}\;,\qquad
\sin^2\theta_l=\frac{1}{b}\;,\qquad
\sin^2\gamma=\frac{1}{K}\;.
\end{equation}

To succeed perfectly in finding the target block, the following
condition should be satisfied
\begin{equation}
|\langle \psi_{\rm rem} | G_g G^{j_l}_l G^{j_g}_g|\psi_{\rm
init}\rangle| = 0\;,
\end{equation}
where $|\psi_{\rm rem}\rangle$ is the uniform superposition of all states
outside the target block.
This reduces to the number of global number $j_g$ and local number
$j_l$ of Grover iterations as satisfying \cite{KorepinLiao}
\begin{eqnarray}
\cos (2j_l\theta_l)&=&\frac{\tan \gamma}{ \tan 2\gamma}
=\frac{K-2}{2(K-1)} \nonumber, \\
\tan (2j_g\theta_g)&=& \frac{\cos 2\gamma }{\sin \gamma\, \sqrt{3-4
\sin^2 \gamma} }
=\frac{K-2}{\sqrt{3K-4}}
\;. \label{korepin}
\end{eqnarray}

Unfortunately, in general $j_l$ and $j_g$ satisfying these equations
are not integers. Hence, in the GRK algorithm in order to minimize
error in the final state one instead should use
$\hat{j_l}=\lfloor{j_l}\rceil$ and $\hat{j_g}=\lfloor{j_g}\rceil$
number of local and global iterations respectively (where
$\lfloor{r}\rceil$ is the nearest integer to $r$). Naturally, this
approximation causes some error in the partial search. We shall now
show how to modify the GRK algorithm to guarantee sure success, thus
overcoming this difficulty. For the moment, however, we shall leave
the the number of local and global iterations $j_l$ and $j_g$ as
unspecified.

\section{Phase Condition for Sure Success}

The special case of sure success partial search is the sure success
full search --- in other words, a sure success variation of the
usual Grover search algorithm. Many approaches to guaranteeing the
ideal behavior of full search have been proposed
\cite{HOYER,LONG-LI-SUN,2000quant.ph..5055B}. In this work, however,
we shall investigate a variation of what we consider to be the
simplest of these, given originally by Brassard {\it et al.\/}
\cite{2000quant.ph..5055B}, which only requires modifying the final
global operator iteration of the entire algorithm.

Firstly, as in Ref.~\onlinecite{KorepinLiao}, we note that the
entire action of the partial search may be compactly described by a
3-dimensional subspace spanned by the vectors: $|x_{\rm sol}\rangle$
the unique solution to the full search; $|\psi_{\rm target}'\rangle$
the normalized target block state {\it excluding\/} the solution
state; and $|\psi_{\rm rem}\rangle$ for other states. Using these
three basis, the initial state $|\psi_{\rm init}\rangle$ of
Eq.~(\ref{init}) may be written
\begin{eqnarray}
|\psi_{\rm init}\rangle&=&\sin\gamma\sin\theta_l|x_{\rm
sol}\rangle+\sin\gamma \cos\theta_l|\psi_{\rm
target}'\rangle\nonumber \\
&&+\cos\gamma|\psi_{\rm rem}\rangle\;.
\end{eqnarray}

Similarly, the ideal target state $|\psi_{\rm target}\rangle$ of
Eq.~(\ref{target}) is
\begin{equation}
|\psi_{\rm target}\rangle=\sin\theta_l|x_{\rm
sol}\rangle+\cos\theta_l|\psi_{\rm target}'\rangle\;.
\end{equation}
The $G^{j_g}_g$ operator is represented as
\begin{equation}
G^{j_g}_g= TM_{j_g}T,
\end{equation}

where
\begin{equation}
T= \left(
\begin{array}{ccc}
1&0&0\\
0&\frac{\cos\theta_l \sin\gamma}{\cos\theta_g}&\frac{\cos\gamma}{\cos\theta_g} \\
0&\frac{\cos\gamma}{\cos\theta_g}&-\frac{\cos\theta_l \sin\gamma}{\cos\theta_g}
\end{array} \right)
\end{equation}
and

\begin{equation}
M_{j_g}= \left(
\begin{array}{ccc}
\cos(2j_g\theta_g)&\sin(2j_g\theta_g)&0\\
-\sin(2j_g\theta_g)&\cos(2j_g\theta_g)&0\\
0&0&(-1)^{j_g}
\end{array} \right)\;.
\end{equation}

The $G^{j_l}_l$ operator is represented as \cite{KorepinLiao}
\begin{equation}
G^{j_l}_l=\left(
\begin{array}{ccc}
\cos(2j_l\theta_l)  &   \sin(2j_l\theta_l)  &   0 \\
-\sin(2j_l\theta_l) &   \cos (2j_l\theta_l) &   0 \\
0                   &                   0   &   1
\end{array}
\right)\;.
\end{equation}

\begin{widetext}
The intermediate state after $j_g$ global Grover iterations is given
by \cite{KorepinLiao}
\begin{eqnarray}
G_g^{j_g}|\psi_{\rm init}\rangle=
\frac{1}{\cos^2\theta_g}
\left(
\begin{array}{c}\cos\theta_g
\left( s_g m + c_g\cos\theta_g
\sin\theta_g\right) \\
\cos\theta_l\sin\gamma\left(c_g m-s_g\cos\theta_g \sin\theta_g
\right)\\
\cos\gamma\left(c_g m-s_g \cos\theta_g \sin\theta_g \right)
\end{array}
\right) \!,
\end{eqnarray}
where $c_g=\cos(2j_g\theta_g)$, $s_g=\sin(2j_g\theta_g)$ and
$m=\cos^2\theta_l\sin^2\gamma+\cos^2\gamma$.

The next intermediate state after $j_g$ global and $j_l$ local
Grover iterations is given by \cite{KorepinLiao}
\begin{eqnarray}
G_l^{j_l}G_g^{j_g}|\psi_{\rm init}\rangle=
\frac{1}{\cos^2\theta_g}
\left(
\begin{array}{c}
c_l\cos\theta_g\left( s_g m+c_g\cos\theta_g \sin\theta_g\right)+
s_l\cos\theta_l\sin\gamma\left(c_g m-s_g\cos\theta_g\sin\theta_g
\right) \\
-s_l\cos\theta_g\left(s_g m+c_g\cos\theta_g\sin\theta_g\right)+
c_l\cos\theta_l\sin\gamma\left(c_g m-s_g\cos\theta_g\sin\theta_g
\right)\\
\cos\gamma\left(c_g m -s_g\cos\theta_g \sin\theta_g \right)
\end{array} \right)=\left(\begin{array}{c}a\\ b\\ c\end{array}\right) \!,
\label{14}
\end{eqnarray}
where $c_l=\cos(2j_l\theta_l)$ and $s_l=\sin(2j_l\theta_l)$.

The final global Grover operator iteration is modified with two
phases as in the exact Grover search \cite{2000quant.ph..5055B}
\begin{eqnarray}
G_g^{\rm final}&\equiv&
-\bigl[\openone-(\openone-e^{2i\theta})|\psi_{\rm init}
\rangle\langle \psi_{\rm init}|\bigr]\nonumber \\
&&\times \bigl[\openone-(\openone-e^{i(\phi-\theta)})
|x_{\rm sol}\rangle\langle x_{\rm sol}|\bigr]\;.
\end{eqnarray}

Translating this into the three basis states supporting the entire
computation we obtain
\begin{equation}
G_g^{\rm final}= \left(
    \begin{array}{ccc}
        -e^{i(\phi-\theta)}[1-(1-e^{2i\theta})\sin^2\gamma \sin^2\theta_l]
        & (1-e^{2i\theta})\sin^2\gamma\sin\theta_l\cos\theta_l
        & (1-e^{2i\theta})\cos\gamma\sin\gamma\sin\theta_l\\
        e^{i(\phi-\theta)}(1-e^{2i\theta})\sin^2\gamma
         \sin\theta_l \cos\theta_l
        & (1-e^{2i\theta})\sin^2\gamma \cos^2\theta_l-1
        & (1-e^{2i\theta})\cos\gamma\sin\gamma\cos\theta_l\\
        e^{i(\phi-\theta)}(1-e^{2i\theta})\sin\gamma \sin\theta_l \cos\gamma
        & (1-e^{2i\theta})\sin\gamma \cos\gamma \cos\theta_l
        & (1-e^{2i\theta})\cos^2\gamma-1\\
    \end{array}
\right).
\end{equation}
\end{widetext}

Finally then, our aim of a sure success partial search will be
achieved if we can find two phases, $\theta$ and $\phi$, for the
above final global Grover operator which satisfies the condition
\begin{equation}
|\langle \psi_{\rm rem}|G_g^{\rm final}G_l^{j_l} G_g^{j_g}
|\psi_{\rm init}\rangle| = 0 \;.
\end{equation}

The relevant phase condition then reduces to
\begin{eqnarray}
&&a e^{i(\phi-\theta)}(1-e^{2i\theta})\sin\gamma \sin\theta_l
\cos\gamma\nonumber \\
&+&b(1-e^{2i\theta})\sin\gamma \cos\gamma
\cos\theta_l\nonumber \\
&+&c[(1-e^{2i\theta})\cos^2\gamma-1]= 0\;, \label{phaseCond}
\end{eqnarray}
where $a$, $b$ and $c$ are defined in Eq.~(\ref{14}).

The phase condition may then be rewritten as
\begin{equation}
e^{i(\phi-\theta)}(1-e^{2i\theta}) x +(1-e^{2i\theta})y+2z = 0\;,
\label{phase_cond_2}
\end{equation}
where
\begin{eqnarray}
x&\equiv&a\sin\gamma \sin\theta_l
\cos\gamma\nonumber, \\
y&\equiv&b\sin\gamma \cos\gamma \cos\theta_l + c\cos^2\gamma
\nonumber,\\
z&\equiv&-\frac{c}{2}\label{xyz} \;.
\end{eqnarray}
The real and imaginary parts of Eq.~(\ref{phase_cond_2}) may be
simplified to give
\begin{eqnarray}
\sin\phi &=&
-\frac{y}{x}\sin\theta  -\frac{z}{x\sin\theta } \nonumber, \\
\cos\phi&=& -\frac{y}{x}\cos\theta \;.
\label{cos}
\end{eqnarray}
Finally, combining these two equations together, we may eliminate $\phi$
to yield
\begin{equation}
\sin^2\theta = \frac{z^2}{x^2-y^2-2yz} \label{sin2} \;,
\end{equation}
which to have a solution must satisfy
\begin{equation}
x^2 \ge (y+z)^2 \label{ineq} \;.
\end{equation}
There will then be a solution for $\phi$ provided the right-hand-sides of
Eq.~(\ref{cos}) are bounded in absolute value by unity.

\section{Numerical Analysis}

In this section, we explain our numerical results showing that the
phase condition~(\ref{phaseCond}) may be easily solved numerically.
Now the number of local and global search steps must be integers so
in the GRK algorithm \cite{Korepin} their values are chosen to be
$\hat{j_l}=\lfloor j_l \rceil$ and $\hat{j_g}=\lfloor j_g \rceil$.
This works fine for GRK under the assumption that both $b$ and $K$
are large. However, we have sought a more general solution. Using a
numerical search, we found that $\theta$ and $\phi$ could always be
found to satisfy the phase condition~(\ref{phaseCond}) provided we
chose
\begin{eqnarray}
\hat{j_l}&=&\lfloor j_l \rfloor, \nonumber \\
\hat{j_g}&=&\lfloor j_g \rfloor +\{0,1,2\}\;,
\end{eqnarray}
where the floor of both $j_l$ and $j_g$ are chosen, however, the
latter may require one or possibly two extra steps (which we denote
by $\{0,1,2\}$). This strategy was found to work numerically for
$N=Kb \le10^6$ in all cases except for the case $K=2$ and $b=2$. In
practice, finding the parameters to work with would be straight-forward
to implement since for these three options for $j_g$ we may determine
the auxiliary quantities $x$, $y$ and $z$ immediately with
Eq.~(\ref{xyz}) after which $\theta$ and $\phi$ are determined separately
through Eqs.~(\ref{cos}) and~(\ref{sin2}).

\section{Conclusion}

In this work, we have investigated the necessary phase conditions to
guarantee sure success of the partial search algorithm. All the search
action goes on within a three-dimensional Hilbert space and solutions
to the phase conditions may be found very easily.

In principle, since there are many potential generalizations of the
full Grover search algorithm, partial search can also be extended
for more general cases. For example, one could consider extending
the partial search algorithm to the more general case involving,
say, multiple target blocks. Other algorithms based on the Grover
search algorithm such as the algorithm associated with the Boolean
weight decision problem \cite{quant-ph/0410043} could be converted
into, so called, a block weight decision problem.

\section*{Acknowledgement}
BSC was partly supported by the Ministry of Information \& Communication
of South Korea (IT Overseas National Scholarship Program). SLB currently
holds a Royal Society --- Wolfson Research Merit Award.
This research is part of the QIP IRC (www.qipirc.org) supported by
EPSRC (GR/S82176/01).


\begin{thebibliography}{99}                                                                                               %
\bibitem{CS-0900.81019MA}
D.\ Deutsch,
\newblock {Quantum theory, the Church-Turing principle and the
universal quantum computer.}
\newblock {\em Proc. R. Soc. Lond., Ser. A} \textbf{400}(1818):97--117, 1985.

\bibitem{qDJ92}
D.\ Deutsch and R.\ Jozsa,
\newblock {Rapid solution of problems by quantum computation.}
\newblock {\em Proc. R. Soc. Lond., Ser. A} \textbf{439}(1907):553--558, 1992.

\bibitem{SICOMP::Shor1997}
P.\ W.\ Shor,
\newblock Polynomial-time algorithms for prime factorization and discrete
logarithms on a quantum computer.
\newblock {\em SIAM Journal on Computing}
\textbf{26}(5):1484--1509, October 1997.

\bibitem{qGR96}
L.\ Grover,
\newblock A fast quantum mechanical algorithm for database search.
\newblock In {\em Proceedings of 28th Annual Symposium on the Theory
of Computing (STOC)}, May 1996, pages 212--219. {\em quant-ph/9605043}.

\bibitem{KorepinGrover}
V.\ E.\ Korepin and L.\ K.\ Grover,
\newblock   {Simple Algorithm for Partial Quantum Search}
\newblock   {\em Quantum Information Processing}
\textbf{5}(1), 3, 2006.
{\em    quant-ph/0504157}

\bibitem{GroverRadhakrishnan}
L.\ K.\ Grover and J.\ Radhakrishnan,
\newblock   {Is partial quantum search of a database any easier?}
\newblock   {\em  Proceedings of the 17th annual ACM Symposium on
Parallelism in Algorithms and Architectures}, July 2005, pages
186--194. {\em quant-ph/0407122}

\bibitem{Korepin}
V.\ E.\ Korepin,
\newblock   {Optimization of partial search}
\newblock   {\em    J.Phys.A:Math. Gen. 38(2005) L731-758}.

\bibitem{KorepinLiao}
V.\ E.\ Korepin and J.\ Liao,
\newblock   {Quest for Fast Partial Search Algorithm}
\newblock   {\em    quant-ph/0510179}, October 2005.

\bibitem{HOYER}
P.\ H{\o}yer,
\newblock {Arbitary Phases in Quantum Amplitude Amplification}.
\newblock {\em Phys. Rev. A} \textbf{62} 052304, 2000.

\bibitem{LONG-LI-SUN}
Y.\ Sun, G.-L.\ Long and X.\ Li,
\newblock {Phase Matching Condition for Quantum Search with a
Generalized Initial State}.
\newblock {\em Phys. Let. A} \textbf{294} 143, 2002.

\bibitem{2000quant.ph..5055B}
G.\ {Brassard}, P.\ {H{\o}yer}, M.\ {Mosca} and A.\ {Tapp},
\newblock {Quantum Amplitude Amplification and Estimation}.
\newblock {\em quant-ph/0005055}, May 2000.
\newblock   {Quantum Computation \& Information, AMS, Contemporary
Mathematics Series Millenium Volume, edited by Samuel J.\ Lomonaco,
vol.\ 305, pp.53-74.2002}

\bibitem{quant-ph/0410043}
S.\ L.\ Braunstein, B.-S.\ Choi, S.\ Maitra, D.\ Chakrabarti, S.\
Ghosh and P.\ Mukhopadhyay,
\newblock   {Quantum algorithm to distinguish Boolean functions of
different weights}
\newblock   {\em    quant-ph/0410043}.
\end{thebibliography}
\end{document}